\documentclass[twocolumn,showpacs,amsmath,amssymb,pra,graphics,graphicx]{revtex4-1}
       \usepackage{graphicx}
        \usepackage{graphics}
\usepackage{bm}
\usepackage{dcolumn}
\begin{document}

\title{Pearl vortices   in anisotropic superconducting films}

\author{V. G. Kogan}
\email{kogan@ameslab.gov}
 \affiliation{Ames Laboratory--DOE, Ames, IA 50011, USA}
  \author{N. Nakagawa}
 \affiliation{Iowa State University, Ames, IA 50011, USA}   
  \author{J. R. Kirtley}
 \affiliation{Kirtleyscientific, Pacific Grove, CA 93950, USA}   
 
  \date{\today}
       
\begin{abstract}
The magnetic field of  vortices in anisotropic superconducting films is  considered  in the framework of anisotropic  London approach. It is found that at distances large relative to the core size, the magnetic field normal to the film surface may change sign.  
  We find that  the magnetic field attenuates at large distances as $1/r^3$ as it does in isotropic films, but the anisotropy induces an angular dependence to the supercurrents which causes the sign of the field to change for anisotropy parameters $\gamma=\lambda_2/\lambda_1>\sqrt{2}$ in some parts of the $(x,y)$ plane.
   \end{abstract}

\maketitle
\section{Introduction}

The magnetic field distribution due to a vortex in  thin isotropic superconducting film in the $(x,y)$ plane was evaluated by J. Pearl  \cite{Pearl}.  The major feature of this distribution is that the field above the film is reminiscent of one due to a magnetic ``charge" $\phi_0$ at the vortex core that spreads  into $2\pi$ solid angle of free space as does the electric field of the point charge. At large distances the field component $h_z$ perpendicular to the film at the film face decays as $1/r^3$,  has everywhere the same sign at the film face (positive for the vortex magnetic flux directed along $+z$), and diverges as $ 1/r$ if $ r\ll \Lambda=2\lambda^2/d$ ($\lambda$ is the London penetration depth of the bulk material, $d$ is the film thickness). These features where confirmed experimentally, see e.g. \cite{Tafuri}.  

The interest in films that are anisotropic in-plane was somewhat muted mainly because of difficulties in their preparation. Films of orthorhombic materials like YBCO, relatively easy to grow with the $c$ axis perpendicular to the film plane, have too small $ab$ anisotropy to show substantial differences with isotropic films. 

 However, recently materials with large in-plane anisotropy were discovered.   STM studies of films made of these materials show the vortex core anisotropies  of about $2.5 - 3.5$ \cite{NiBi,17,Ta4Pd3Te16}.  The question then arises whether vortices in films of these materials have the same Pearl form just amended with a proper rescaling. In this work  we show that this is not the case. The magnetic field structure of anisotropic Pearl vortex   differs qualitatively from its isotropic version. In particular, for large enough anisotropy the field crossing the film may change sign in some patches of the $x,y$ plane, nevertheless keeping the total flux associated with vortex equal to the flux quantum $\phi_0$.

\section{Thin films}

We begin with the outline of our approach for isotropic films.    Let a film of thickness $d$ be in the $xy$ plane. Integration of the London equation for the magnetic field, ${\bm h}- \lambda^2\nabla^2{\bm h}  = \phi_0 \hat{\bm z} \delta({\bm r} )$, over the film  thickness gives for the $z$ component of the field at the film:
\begin{eqnarray}
\frac{2\pi\Lambda}{c}{\rm curl}_z {\bm g} + h_z   =\phi_0 \delta(\bm r)    .
\label{2D London}
\end{eqnarray}
Here,  $\bm g$ is the sheet current density related to the tangential field components at the upper film face by  $2\pi\bm g/c=\hat{\bm z}\times \bm h$; $\Lambda=2\lambda^2/d$ is the Pearl length.  With the help of div$\bm h=0$ this equation is transformed to:
\begin{eqnarray}
h_z -\Lambda \frac{\partial h_z}{\partial z}    =\phi_0 \delta(\bm r) .
\label{hz-eq}   
\end{eqnarray}
 
A  large contribution to the energy of a vortex in a thin film comes from  stray fields \cite{Pearl}. The problem of a vortex in a thin film is, in fact, reduced to that of the field distribution in free space subject to the boundary condition supplied by solutions of Eq.\,(\ref{2D London}) at the film surface. Since outside the film curl$\bm h=\,\,\,$div$\bm h=0$, one can introduce a scalar potential for the {\it outside} field in the upper half-space:
 \begin{eqnarray}
\bm h   =\bm \nabla \varphi,\qquad \nabla^2\varphi=0   \,.
\label{define _phi} 
\end{eqnarray}
The general form of the potential satisfying Laplace equation that vanishes at $z\to\infty$   is 
 \begin{eqnarray}
\varphi (\bm r, z)   =\int \frac{d^2\bm k}{4\pi^2} \varphi(\bm k) e^{i\bm k\cdot\bm r-kz}\,.
\label{gen_sol} 
\end{eqnarray}
Here, $\bm k=(k_x,k_y)$, $\bm r=( x, y)$, and  $ \varphi(\bm k)$  is the two-dimensional (2D) Fourier transform of $ \varphi(\bm r, z=0)$. In the lower half-space one has to replace $z\to -z$ in Eq.\,(\ref{gen_sol}). 

One applies now the 2D Fourier transform to Eq.\,(\ref{hz-eq}) to obtain:
\begin{eqnarray}
h_{z\bm k}   =-k\varphi_{\bm k}= \frac{\phi_0  }{ 1+\Lambda k }  \, 
\label{hz(k)} 
\end{eqnarray}
 As mentioned above, the sheet current is related to the tangential field components by
\begin{eqnarray}
\frac{2 \pi}{c}g_x = -h_y\,,\qquad \frac{2 \pi}{c}g_y =  h_x\,.
\label{current1} 
\end{eqnarray}
In 2D Fourier space,    $h_{x\bm k}   =ik_x\varphi_{\bm k}$ and $h_{y\bm k}   =ik_y\varphi_{\bm k}$ and   
 we obtain:
\begin{eqnarray}
 g_{x \bm k} = -g_{y \bm k}\,\frac{k_y}{k_x} = \frac{c\phi_0}{2\pi }  \frac{i k_y }{k(1+k\Lambda)}. \qquad   \label{gx} 
\end{eqnarray}

Thus,  the field $\bm h$ in the free space and at the film along with the currents can be expressed in terms of the potential $\varphi$. 
 It is easy to see that  stream lines of the  current  coincide with  contours of $\varphi(x,y)=\,\,$const.  
Moreover, the self-energy of a Pearl vortex and the interaction energy of two vortices can be expressed in terms of $\varphi$ \cite{BKT}. 

\subsection{Magnetic flux of a Pearl vortex}

Eq.\,(\ref{hz(k)}) gives the Fourier transform of the field $h_z$ at the film surface.
 At a finite hight $z$ above the film:
\begin{eqnarray}
h_{z } (\bm k,z)  =  \frac{\phi_0 e^{-kz} }{ 1+\Lambda k }  \, 
\label{hz(k)1} 
\end{eqnarray}
and  
 \begin{eqnarray}
  h_z(\bm r,z)  = \frac{ \phi_0}{4\pi^2 } \int  \frac{d^2\bm k\, e^{i\bm k \bm r - 
  kz} }{ 1+\Lambda k }\,.
    \label{h(r,z)}
\end{eqnarray}
As described in Appendix A, this can be transformed  to 
  \begin{eqnarray}
 h_z(  R,Z)  = \frac{\phi_0}{2\pi \Lambda^2}   \int_0^\infty du \frac{e^{-u}(u+Z)}{[R^2+(u+Z)^ 2]^{3/2}} \,,\qquad
   \label{dh2}
\end{eqnarray}
where $R=r/\Lambda$ and $Z=z/\Lambda$.  

\begin{figure}[t]
\includegraphics[width=7.5cm] {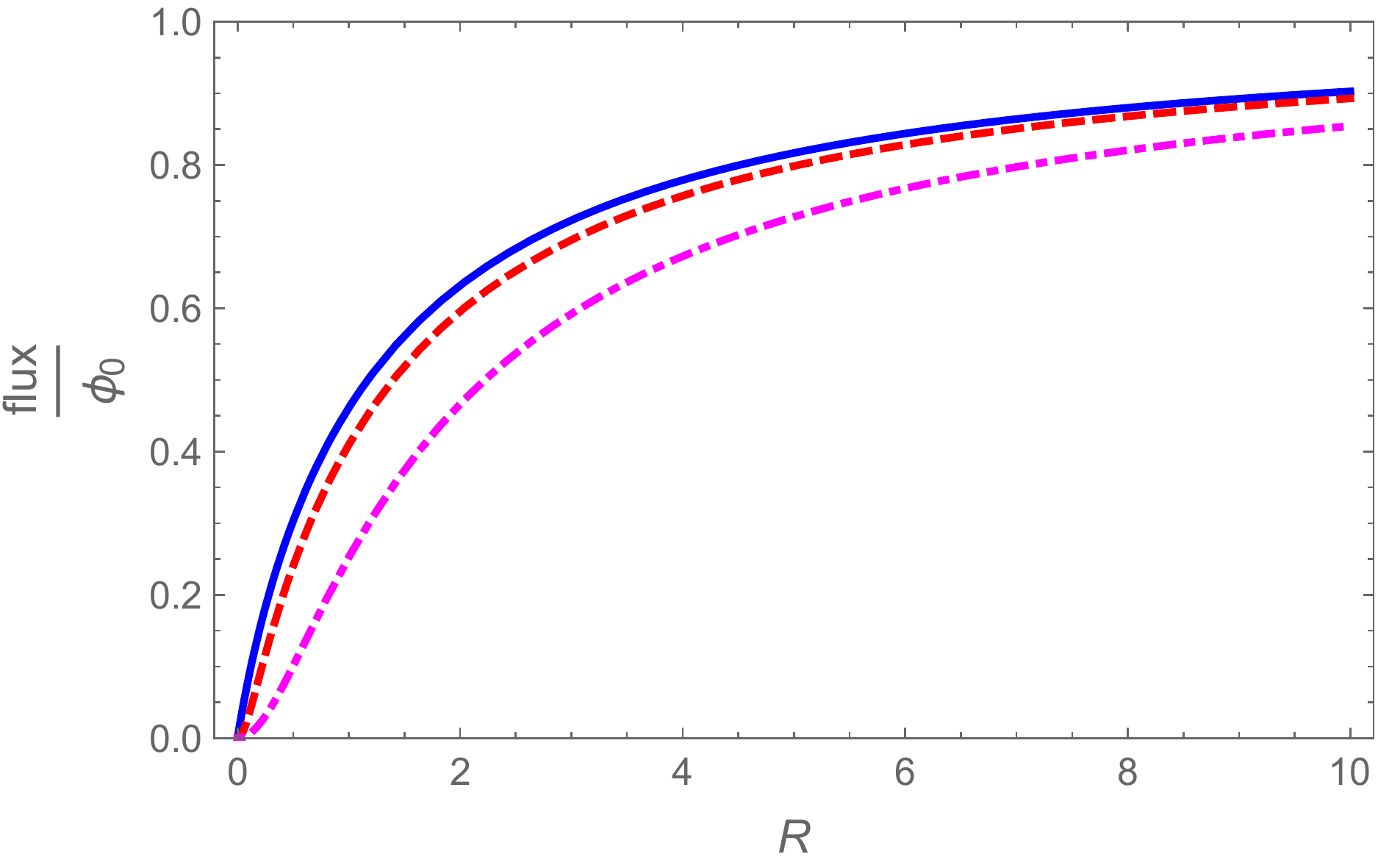}
\caption{ The blue line is $\Phi_z(  R,Z)$ for $Z=0$, the dashed-red is for $Z=0.1$, and the magenta dot-dashed is for $Z=0.5$.}
\label{f1}
\end{figure}

Now, one can calculate the flux through a circle of the radius $R$:
  \begin{eqnarray}
 \Phi_z(  R,Z)  &=&  \Lambda^2\int_0^R 2\pi   R\,dR\, h_z(R,Z)\nonumber\\
 &=& \phi_0\left(1-\int_0^\infty \frac{du\,e^{-u}(u+Z)}{\sqrt{R^2+(u+Z)^ 2}}\right)\,.
   \label{Fz(R)}
\end{eqnarray}
For $R\to\infty$, one can replace the denominator here by $R$ to obtain
  \begin{eqnarray}
 \phi_0 -\Phi_z(  R,Z)  \sim \phi_0\frac{1+Z}{R}  \,.
   \label{Fz(R>>1)}
\end{eqnarray}
Hence,   $\Phi_z$ approaches $\phi_0$ very slowly  as $1/R$.
The flux $\Phi_z(R)$ is plotted in Fig.\,\ref{f1} for $Z=0,\,\,0.1,\,\,0.5$.
It is seen that even for $r=10\,\Lambda$, the flux   reaches only $\sim 0.8\phi_0$.  Note for comparison that in the bulk the flux $\Phi_z \approx 0.9998$ at $r=10\,\lambda$.
  
\section{Anisotropic  films}

The London equations for an arbitrary oriented vortex in anisotropic material  have been given in \cite{K81}. In general, results are cumbersome, so here we consider a simple situation of an orthorhombic superconductor in  field   along the $c$ axis.  The  London equation for the vortex along $z$ in the bulk is:
 \begin{eqnarray}
  h_z +\frac{4\pi}{c} \left(\lambda^2_{yy} \, \frac{\partial J_y}{\partial x}- \lambda^2_{xx} \, \frac{\partial J_x}{\partial y}\right) = \phi_0\delta(\bm r)\,,
 \label{hz} 
 \end{eqnarray} 
 Here, the frame $x,y,z$ is chosen to coincide with $a,b,c$ of the crystal, $\bm r=(x,y)$,   $\lambda^2_{xx} $ and $\lambda^2_{yy} $ are the diagonal components of the tensor $(\lambda^2)_{ik} $. A thin film of this material is assumed to be in the $(x,y)$ plane. Integrating this over the film thickness one obtains:
  \begin{eqnarray}
  h_z +\frac{2\pi}{c} \left(\Lambda_{yy} \, \frac{\partial g_y}{\partial x}- \Lambda_{xx} \, \frac{\partial g_x}{\partial y}\right) = \phi_0\delta(\bm r)\,,
 \label{hz-film} 
 \end{eqnarray} 
where $\bm g=\bm J d$ is the sheet current and 
 $\Lambda_{xx}=2\lambda_{xx}^2/d=\Lambda_1$ and  $\Lambda_{yy}=2\lambda_{yy}^2/d=\Lambda_2$ are the principal Pearl lengths.  
 Taking into account Eq.\,(\ref{current1}) we obtain for the field components  at the film surface:
  \begin{eqnarray}
  h_z +  \Lambda_2 \, \frac{\partial h_x}{\partial x}+\Lambda_1 \, \frac{\partial h_y}{\partial y}  = \phi_0\delta(\bm r)\,.
 \label{hz-film} 
 \end{eqnarray} 
 Since $h_{x\bm k}=ik_x\varphi_{\bm k}$, $h_{y\bm k}=ik_y\varphi_{\bm k}$, and $h_{z\bm k}=-k\varphi_{\bm k}$,
 the 2D FT yields the potential \cite{KSimLed}:
  \begin{eqnarray}
 \varphi_{\bm k} =- \frac{\phi_0 }{ k+\Lambda_1 k_y^2+\Lambda_2 k_x^2 } \,.
\label{phi(k)} 
\end{eqnarray}
 Introduce now the anisotropy parameter $\gamma^2=\Lambda_2/\Lambda_1$ and  $\Lambda=\sqrt{\Lambda_1\Lambda_2}$ so that $\Lambda_2=\Lambda \gamma$ and $\Lambda_1=\Lambda/ \gamma$ and take $\Lambda$   as the unit length:
  \begin{eqnarray}
 h_{z\bm q} = -q \varphi_{\bm q}=\frac{\phi_0 q}{ q+q_x^2  \gamma +q_y^2 /\gamma } \,,\quad \bm q=\bm k \Lambda.
\label{hz,phi} 
\end{eqnarray}

\subsection{Distribution of $\bm{h}_z$} 

At a finite hight $z$ above the film:
\begin{eqnarray}
h_{z } (\bm q,z)  =  \frac{\phi_0\,q\, e^{-qZ} }{ q+q_x^2 \gamma +q_y^2/\gamma }  \, .
\label{hz(k,z)1} 
\end{eqnarray}
Hence, 
 \begin{eqnarray}
  h_z(\bm R,Z)  = \frac{ \phi_0}{4\pi^2 } \int  \frac{d^2\bm q\,q\, e^{i\bm q \bm R - 
  qZ} }{ q+q_x^2 \gamma +q_y^2/\gamma  },
    \label{h(R,Z)}
\end{eqnarray}
where $\bm R=\bm r/\lambda$ and $Z=z/\Lambda$.

Let us consider the field $h_z$ at the film surface, i.e.  $Z=0$. There are two possibilities of dealing with the  integral (\ref{h(R,Z)}). The first is a ``brute force" 2D Fast Fourier Transform (FFT). The second is   to reduce--if possible--the 2D integral over $\bm q$ to a single integration which would be amenable for  numerical evaluation. The second possibility is described in Appendix B with the result:
 \begin{eqnarray}
&& h_z \frac{2\pi \Lambda^2}{\phi_0} = 
 -\frac{1}{2} \int_0^\infty  \frac{ d\eta}{\sqrt{\mu\nu} }   \left(\frac{1} {\rho } -\frac{\eta}{2}\right) 
 e^{-\eta\rho/2}  \,,  \nonumber\\
&& 
\mu= 1+\eta\gamma ,\,\,\, \nu= 1+\eta/\gamma,\,\,\, \rho= \sqrt{\frac{X^2}{\mu }+\frac{Y^2}{\nu } } \,.
    \label{Hz1short} 
\end{eqnarray}
 
\begin{figure}[t]
\includegraphics[width=7.5cm] {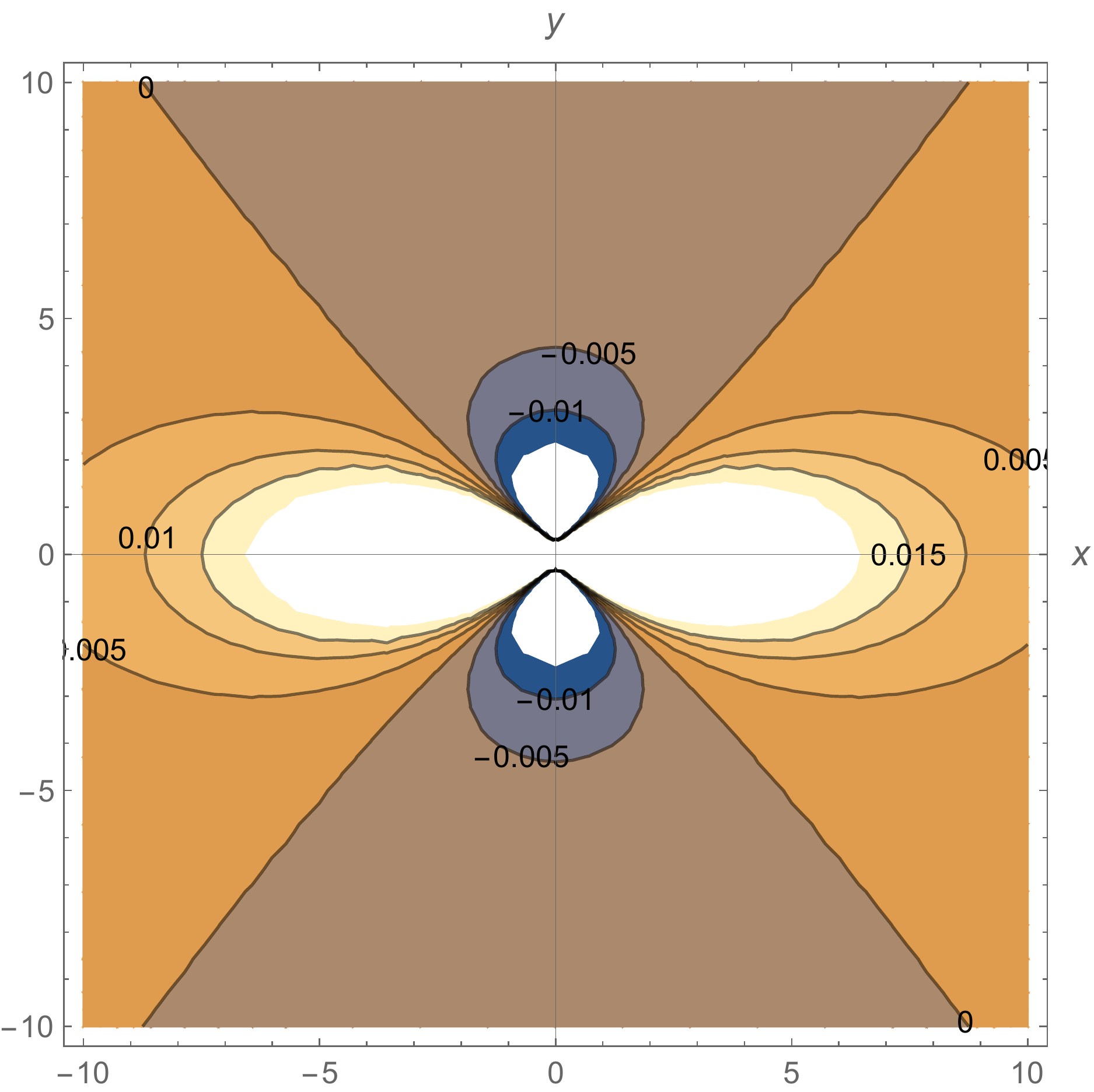}
\caption{ Contours of  $h_z(X,Y)=\,\,$const ($h_z$ is in units  $\phi_0/2\pi\Lambda^2$ and $X,Y$ are in units of $\Lambda$)  for $\gamma=3$.}
\label{f2}
\end{figure}

\begin{figure}[t]
\includegraphics[width=7.5cm] {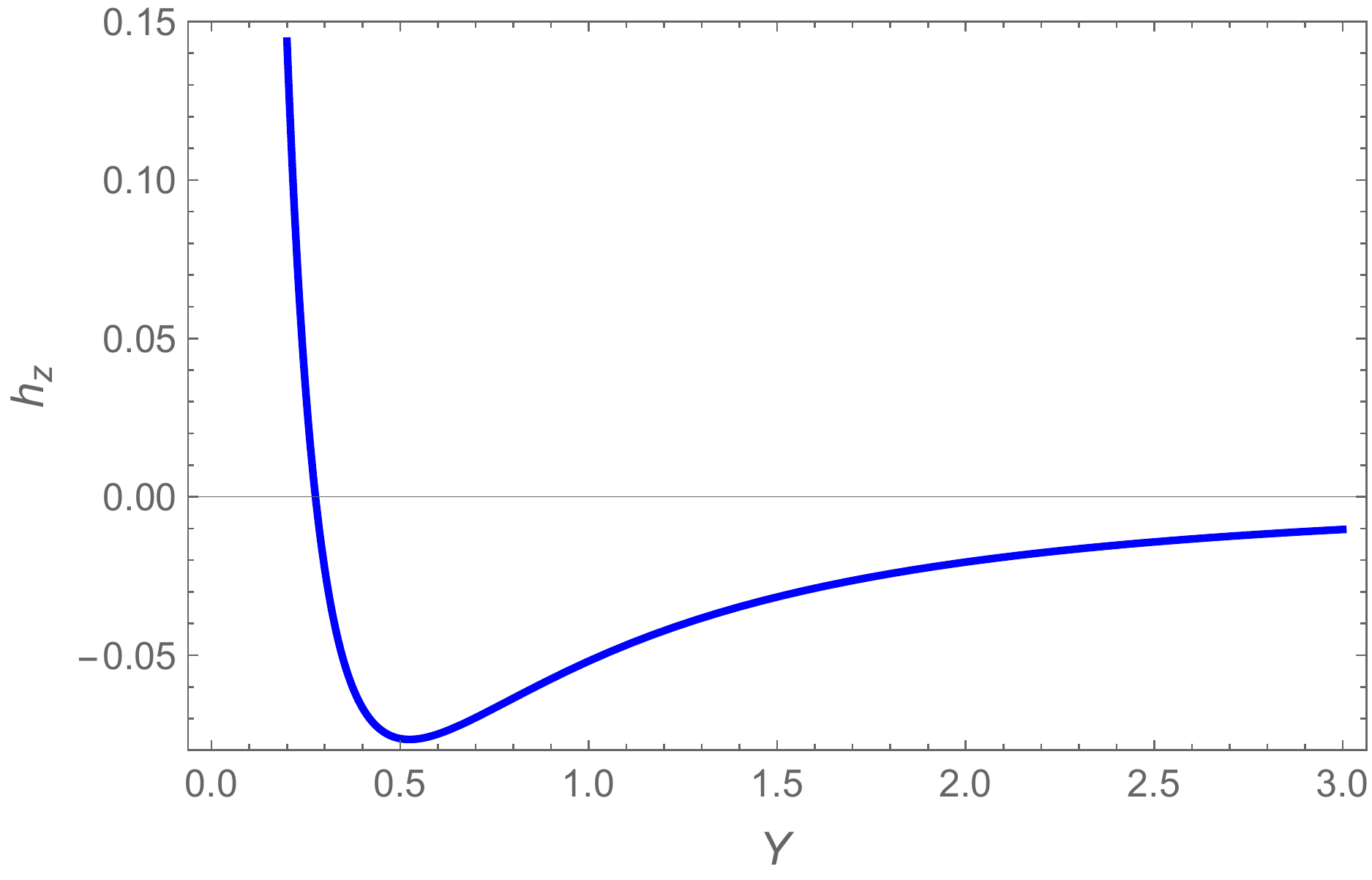}
 \caption{    $h_z(0,Y) $  ($h_z$ is in units  $\phi_0/2\pi\Lambda^2$ and $X,Y$ are in units of $\Lambda$)  for $\gamma=3$.}
\label{f3}
\end{figure}

An example of the field distribution according to this equation is given in Fig.\,\ref{f2}. 
There are two major unexpected features in this result. The first is that the contours $h_z(X,Y)=\,\,$const are not elliptic. 
The second and the most surprising one is that there are  parts of the $X,Y$ plane where the field is negative.
Fig.\,\ref{f3} shows that $h_z(0,Y) $ is positive in vicinity of the singularity at $X=Y=0$, turns zero at $Y\approx 0.3$, changes sign and, after reaching negative minimum near $Y\approx 0.5$, decays to $-0$ as $Y\to\infty$.
It is worth noting that in infinite isotropic film the field $h_z$ crossing the film has the same sign everywhere, i.e., after crossing the film the stray field lines go to infinity never crossing the film again. 
Although the possibility of the field $h_z$ changing sign in anisotropic films was noted time ago \cite{KSimLed}, the phenomenon was not studied in any detail.
  
One can derive analytically  the asymptotic behavior of the field as $R\to\infty$. Omitting details we present the result:
\begin{eqnarray}
 h_z \frac{2\pi \Lambda^2}{\phi_0} &=&  
 \frac{1}{2R^3\gamma}   \left[ \gamma^2+1  +3 (\gamma^2-1) \frac{X^2-Y^2}{R^2} +{\cal O}(R^{-1})\right].\nonumber\\
    \label{as} 
\end{eqnarray}
The second term here does not survive in the isotropic case, whereas for anisotropic films it describes the angular dependent part: $(X^2-Y^2)/R^2=\cos 2\alpha$ where $\alpha$ is the azimuth counted from the $X$ axis. At large distances the lines where $h_z(X,Y)=0$ are given by                               
\begin{eqnarray}
   \cos 2\alpha =-\frac{\gamma^2+1}{3(\gamma^2-1)} \,.
      \label{alpha} 
\end{eqnarray}
 Hence, we obtain a restriction on  $\gamma$ for this solution to exist $\gamma>\sqrt{2}$.
 For $\gamma \gg 1$, $\cos 2\alpha=-1/3$, i.e. $\alpha =54.7^\circ$ and the opening angle of the domain of negative $h_z$ is $\sim 70.6^\circ$.
    \begin{figure}[h]
\includegraphics[width=7cm] {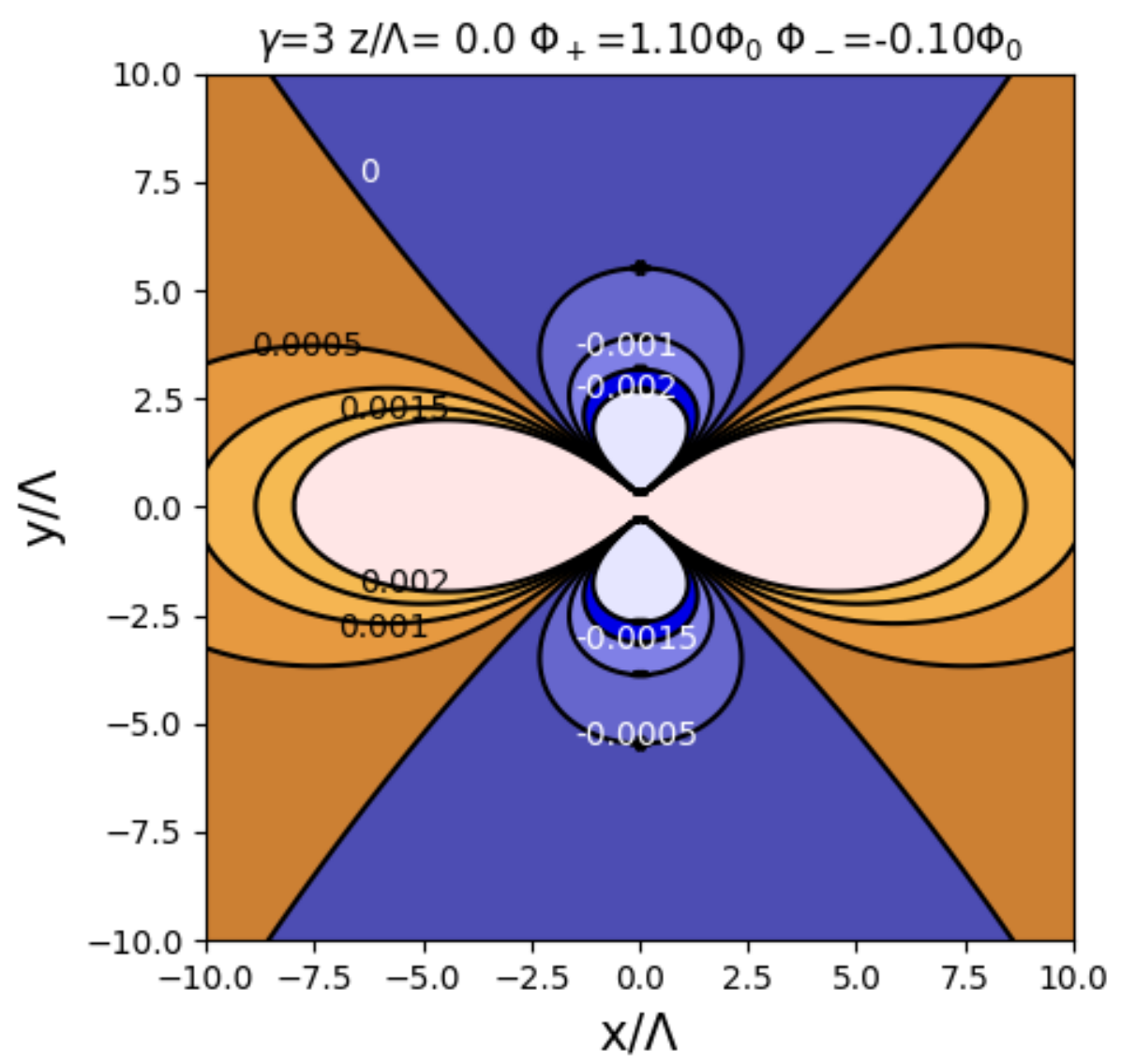}
 \caption{    $h_z(X,Y) $  (in units  $\phi_0/\Lambda^2$ and $X,Y$ in units of $\Lambda$)  for $\gamma=3$ in the window $-10<X,Y<10$ obtained by using $2000\times 2000$ pixel FFT in a cell $-50<X,Y<50$. Here the pink regions (horizontal lobes) have $h_z>0.002$ and the light blue regions (vertical lobes) have $h_z< -0.002$. 
 $\Phi_+ $ is the integrated flux of positive $h_z(X,Y)$ over the entire $50\Lambda \times 50\Lambda$ area and 
 $\Phi_-$ is the integrated flux of negative $h_z(X,Y)$ over the same area.}
\label{f4}
\end{figure}

 Since the profile $h_z(X,Y)$ shown in Fig.\,\ref{f2} is highly unusual and was obtained after an involved analytical procedure, to be confident we applied the 2D FFT directly to the double integral of Eq.\,(\ref{h(R,Z)}). The result shown 
in Fig.\,\ref{f4} confirms existence of the negative domain. Moreover, it allows one to estimate the negative fraction of the flux    as $\sim 10\%$ of the flux quantum $\phi_0$ for $\gamma=3$ at $Z=0$.

\begin{figure}[h]
\includegraphics[width=7cm] {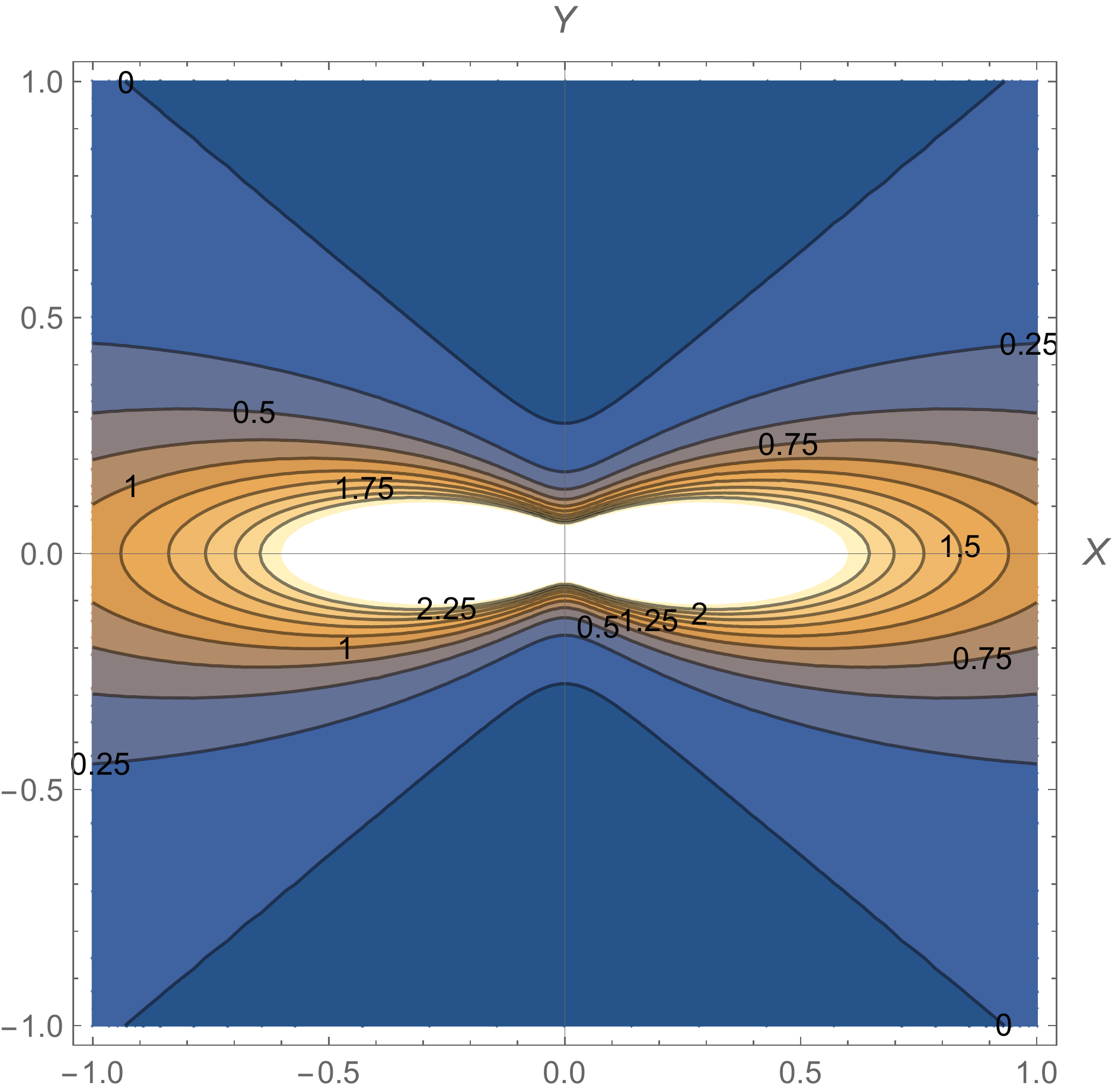}
 \caption{    $h_z(X,Y) $    for $\gamma=3$ for a shorter  distances $-1<(X,Y)<1$. The darkest part is the region of negative $h_z$.}
\label{f5}
\end{figure}

The behavior of $h_z(X,Y)$ at shorter distances $-1 <(X,Y)<1 $ is shown in Fig.\,\ref{f5}.
Note that these distances $\sim\Lambda$ are still large relative to the core size, so that we are still in the region where the London approach holds.

\subsection{$\bm Z$ dependence}

\begin{figure}[htb]
\includegraphics[width=7cm] {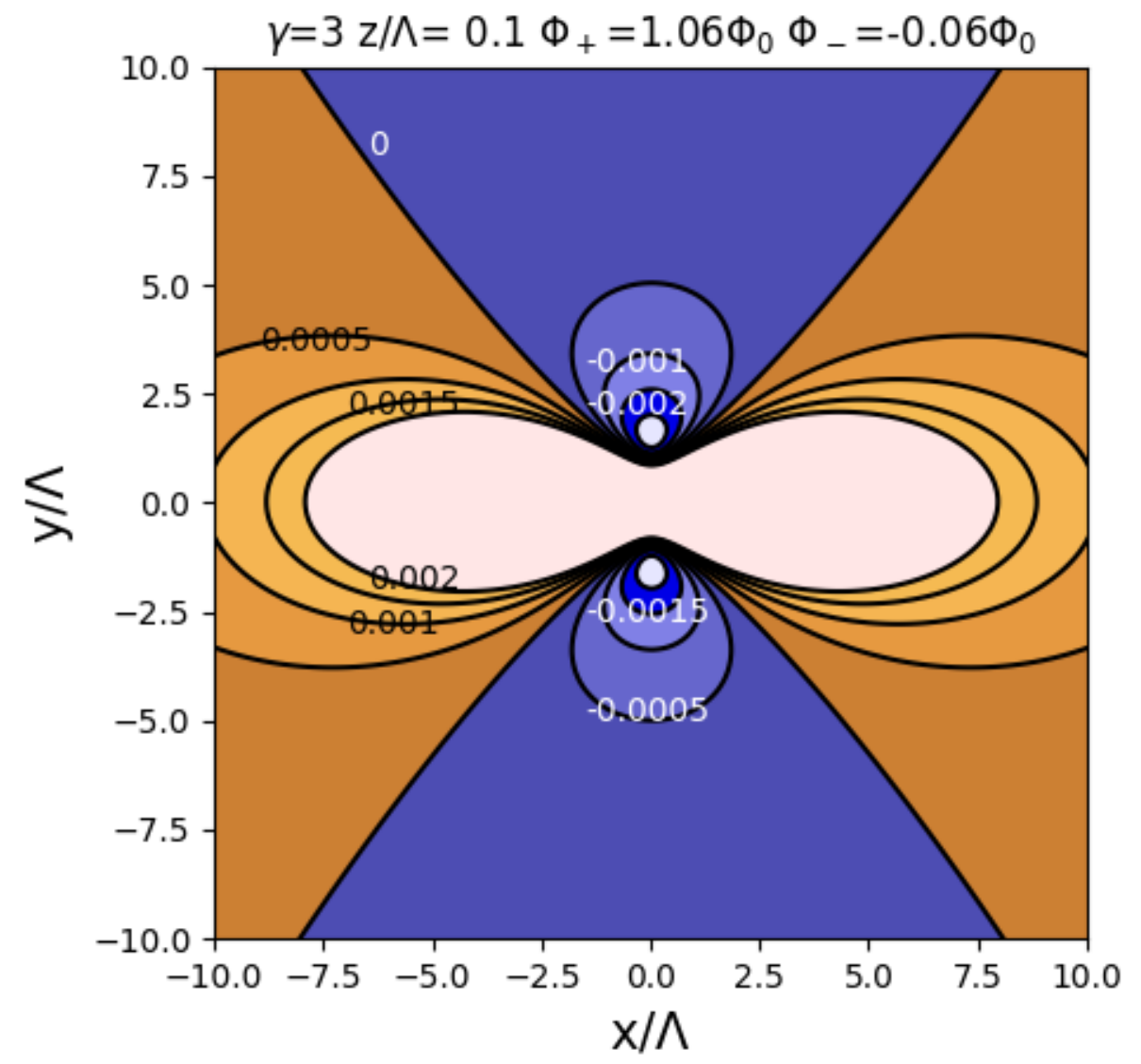}
 \caption{ FFT evaluation of   $h_z(X,Y) $ in units $\phi_0/\Lambda^2$ for $Z=0.1$   done for $-50<(X,Y)<50$, but the result is shown in the window $-10<(X,Y)<10$.}
\label{f6}
\end{figure}

\begin{figure}[htb]
\includegraphics[width=7cm] {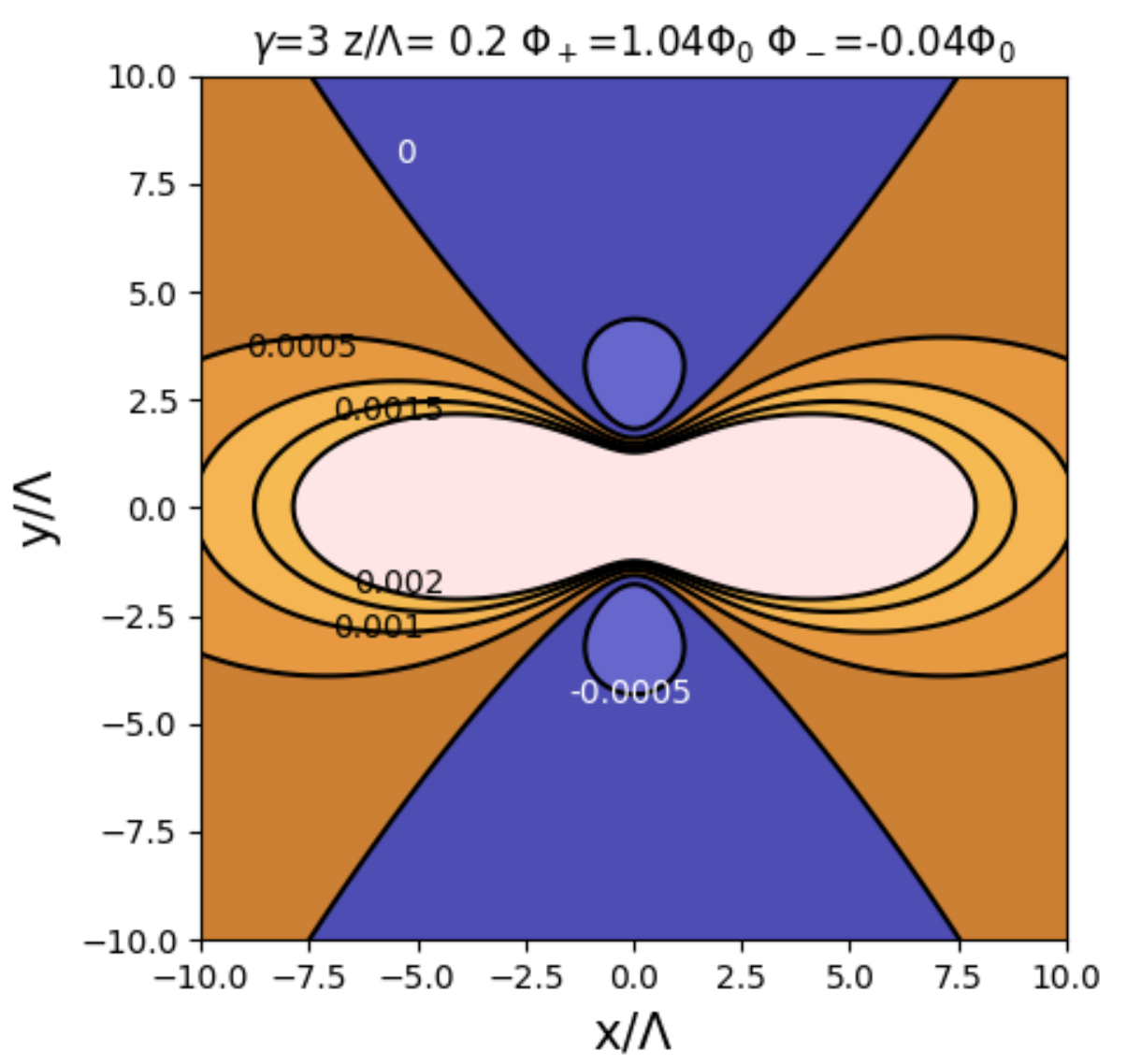} 
 \caption{ FFT evaluation of   $h_z(X,Y) $ in units $\phi_0/\Lambda^2$ for $Z=0.2$   done for $-50<(X,Y)<50$, but the result is shown in the window $-10<(X,Y)<10$  for $Z=0.2$.}
\label{f7}
\end{figure}

\begin{figure}[htb]
\includegraphics[width=7cm] {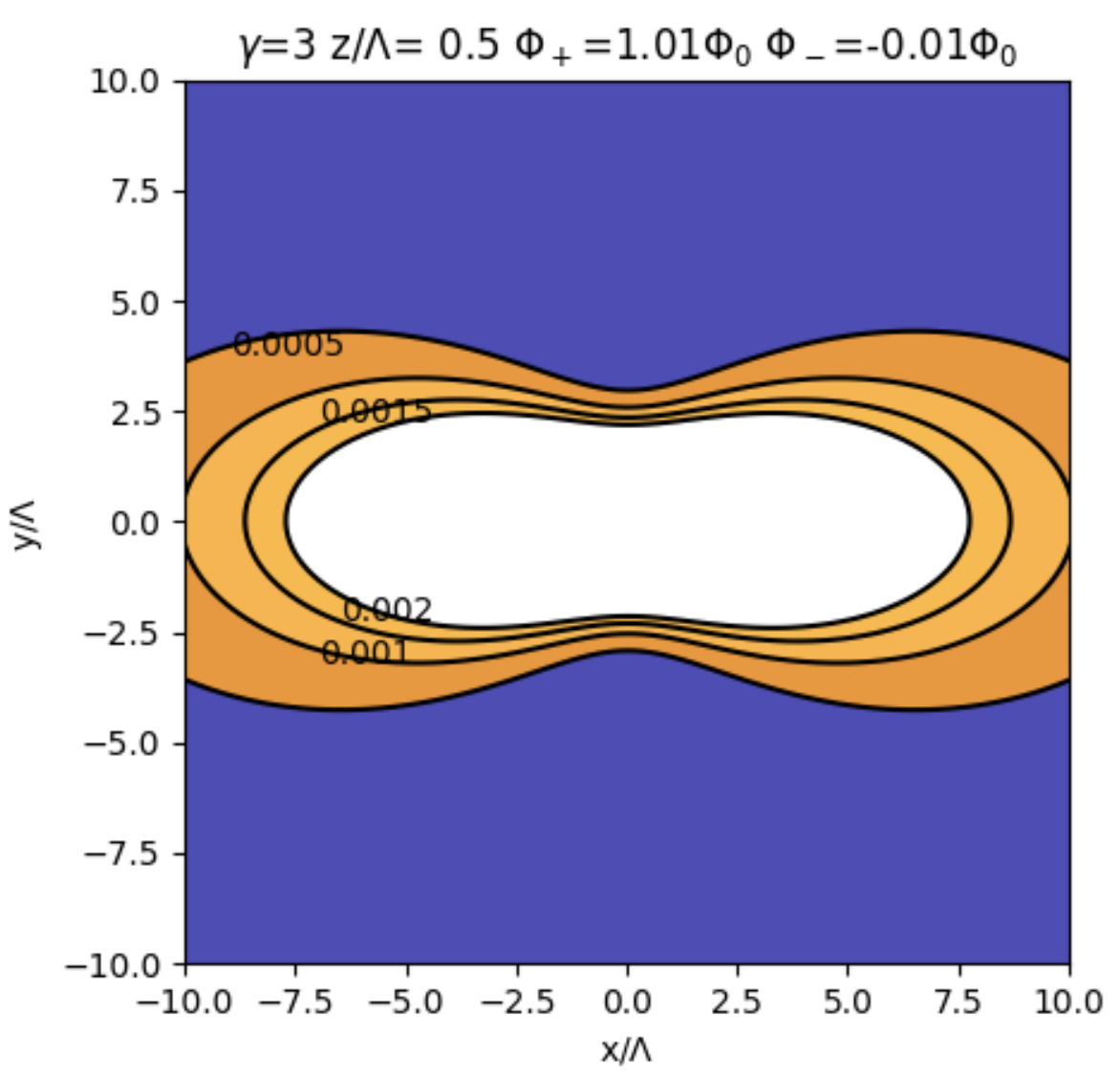}
 \caption{ FFT evaluation of   $h_z(X,Y) $ in units $\phi_0/\Lambda^2$    done for $-50<(X,Y)<50$, but the result is shown in the window $-10<(X,Y)<10$ for $Z=0.5$. }
\label{f8}
\end{figure}

All methods of studying Pearl vortices use sensors placed at a certain height above   the film and measuring $h_z(\bm r,z)$ at a small but finite $z$. Hence, it is of interest to see what Eq.\,(\ref{h(R,Z)}) generates in the case of anisotropic films. The series of $h_z(\bm r,z)$ profiles obtained with the help of FFT is shown in Figs.\,\ref{f6}--\ref{f8}. The negative flux, which is $\approx 0.10\phi_0$ for $Z=0$, decreases to $ 0.06\phi_0$ for $Z=0.1$, to $ 0.04\phi_0$ for $Z=0.2$, and to $ 0.01\phi_0$ for $Z=0.5$. 

In fact, it is possible to get the $h_z(\bm R,Z)$ analytically as a 1D integral in the same manner as Eq.\,(\ref{Hz1short}) was derived:
 \begin{eqnarray}
 h_z \frac{2\pi \Lambda^2}{\phi_0}&=& 
  \frac{1}{2} \int_0^\infty  \frac{ d\eta}{\sqrt{\mu\nu} }   \Big[  \frac{\eta}{2} - \frac{1-\eta Z}{\rho}\nonumber\\
  &+& \frac{Z^2(\eta\rho+2)}{2\rho^3}\Big] \exp\left[-\frac{\eta(\rho+Z)}{2}\right] \,,  \nonumber\\ 
\rho&=& \sqrt{\frac{X^2}{\mu }+\frac{Y^2}{\nu }+Z^2 } \,.
    \label{Hz1lift} 
\end{eqnarray}
where $\mu,\nu$ are given in Eq.\,(\ref{Hz1short}).
 
    \begin{figure}[htb]
\includegraphics[width=8cm] {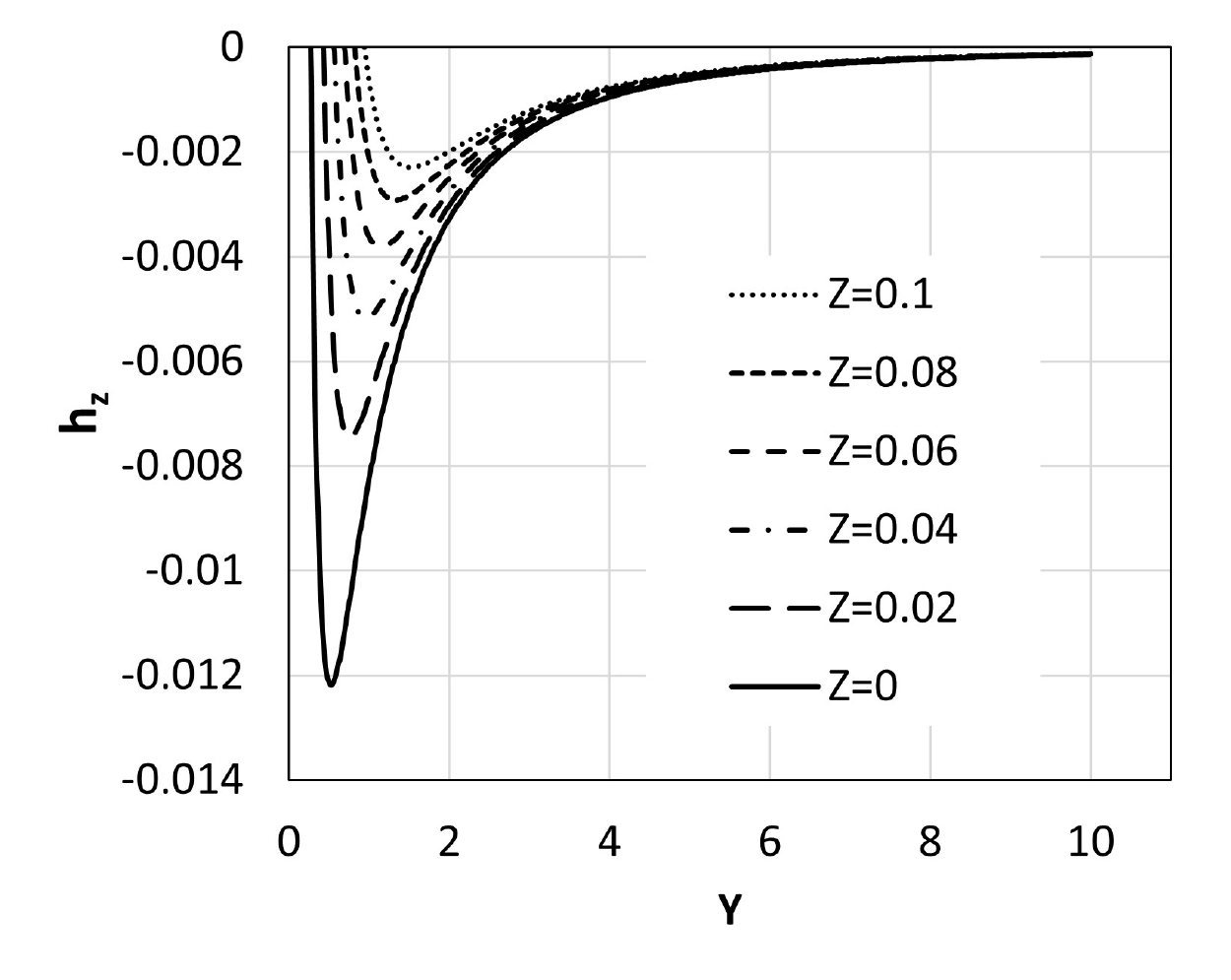}
\includegraphics[width=8cm] {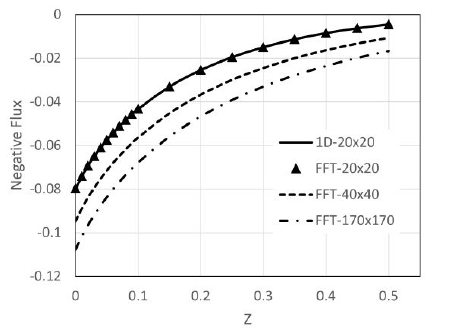}
 \caption{ The upper panel:  $h_z(0,Y) $ in units $\phi_0/2\pi\Lambda^2$ for $Z=z/\Lambda$ indicated in the legend.  The lower panel:  the negative   flux  in units $\phi_0 $ vs the lift $Z $.
 All FFT lines were obtained using results from FFT on a very large  unit cell $170\times 170$. These ``data" were then scanned through smaller windows indicated in the legend. The upper solid curve is obtained evaluating numerically the 1D integral of Eq.\,(\ref{Hz1lift}) for the field distribution and then by integration (summation) over the window area $20\times 20$. The 1D integral  is computed point-wise in the 0.01 steps. 
  Note a good agreement with the FFT results in this window shown by solid triangles.
  }
\label{f9a}
\end{figure}


One can also estimate the negative fraction of the flux $\Phi_z=\int d^2\bm r h_z(\bm r)$. To this end, we did this with the help of both FFT and  employing  $h_z(\bm R,Z)$ of Eq.\,(\ref{Hz1lift}). One should be careful applying FFT to the problem of a single vortex, since within FFT one has to choose a large patch of the $XY$ plane  as a unit cell of periodic lattice covering the whole plane to apply periodic boundary conditions. The hope then is that for large enough unit cell, say, $(50 \times 50)\Lambda$, the field of the vortex is small enough near the cell boundaries where the field   distortions by periodic boundary conditions do not matter.  Doing this one has to   require the flux $\Phi_z$ through the cell $(50 \times 50)\Lambda$ be $\phi_0$. This procedure works well in the bulk where the vortex field decays exponentially and choosing the FFT cell of a few $\lambda$ one gets $\Phi_z$ very close to $\phi_0$. In thin films, however, $(\phi_0-\Phi_z)$ decreases extremely slow as $1/r$ with $r$ being the cell linear size. Hence, to have accurate FFT output one has to choose a large FFT unit cell, e.g.  $(50 \times 50)\Lambda$, and consider only a central part of the cell, say $(10 \times 10)\Lambda$ where the effects of boundaries are weak and the results are  reliable. That is how Figs.\,\ref{f4}, \ref{f6}--\ref{f8} were obtained. 
Repeating the FFT analysis over a larger $(100  \times 100)\Lambda$ area results in positive and negative fluxes $\sim 1\%$ different than reported here for  $(50  \times 50)\Lambda$ area.

Now, the 1D integral representation of $h_z(\bm R,Z)$, Eq.\,(\ref{Hz1lift}), is equivalent to the original 2D integral over $q_x,q_y$, Eq.\,(\ref{h(R,Z)}), in other words, it is a solution of the vortex problem on the infinite $x,y$ plane which satisfies the condition $\Phi_z=\phi_0$. Unlike FFT, we start here with the exact solution, choose, say $(10 \times 10)\Lambda$ window, and   calculate the negative flux  in this window. 

We conclude this section with the plot of $h_z(0,Y)$ along the $Y$ axis  in Fig.\,\ref{f9a} and of integrated negative flux at a set of  heights $Z$.
\subsection{Potential and Currents}

Manipulations, similar to those used for deriving $h_z(\bm r)$, using Eq.\,(\ref{h(R,Z)}) for the potential $\varphi(\bm q)$,  give in real space:
 \begin{eqnarray}
 \varphi(\bm r)  =- \frac{\phi_0 }{4\pi} \int_0^\infty  \frac{d\eta}{ \sqrt{\mu\nu}} \,
 e^{- \eta\rho/2 } \,,
\label{phi(r)1} 
\end{eqnarray}
where $\mu,\nu$ and $\rho$ are defined in Eq.\,(\ref{Hz1short}).

At large distances $r\gg \Lambda$ only small $q$ are relevant, so that the term $q_y^2/\gamma+  q_x^2\gamma $  can be discarded in Eq.\,(\ref{hz,phi}) that implies that the potential there is isotropic. Physically the field there should correspond to that of a point ``magnetic charge" $\phi_0$ in the solid angle $2\pi$ (in the upper half-space), i.e. $\varphi(r)=-\phi_0/2\pi r$. This conclusion could also be reached using  Eq.\,(\ref{phi(r)1}): for $\rho\to\infty$ the relevant $\eta\to 0$ due to the exponential factor. Hence, $\mu \sim \nu\sim 1$ and the remaining integral gives $\varphi(r)=-\phi_0/2\pi r$.

    \begin{figure}[htb]
\includegraphics[width=7.5cm] {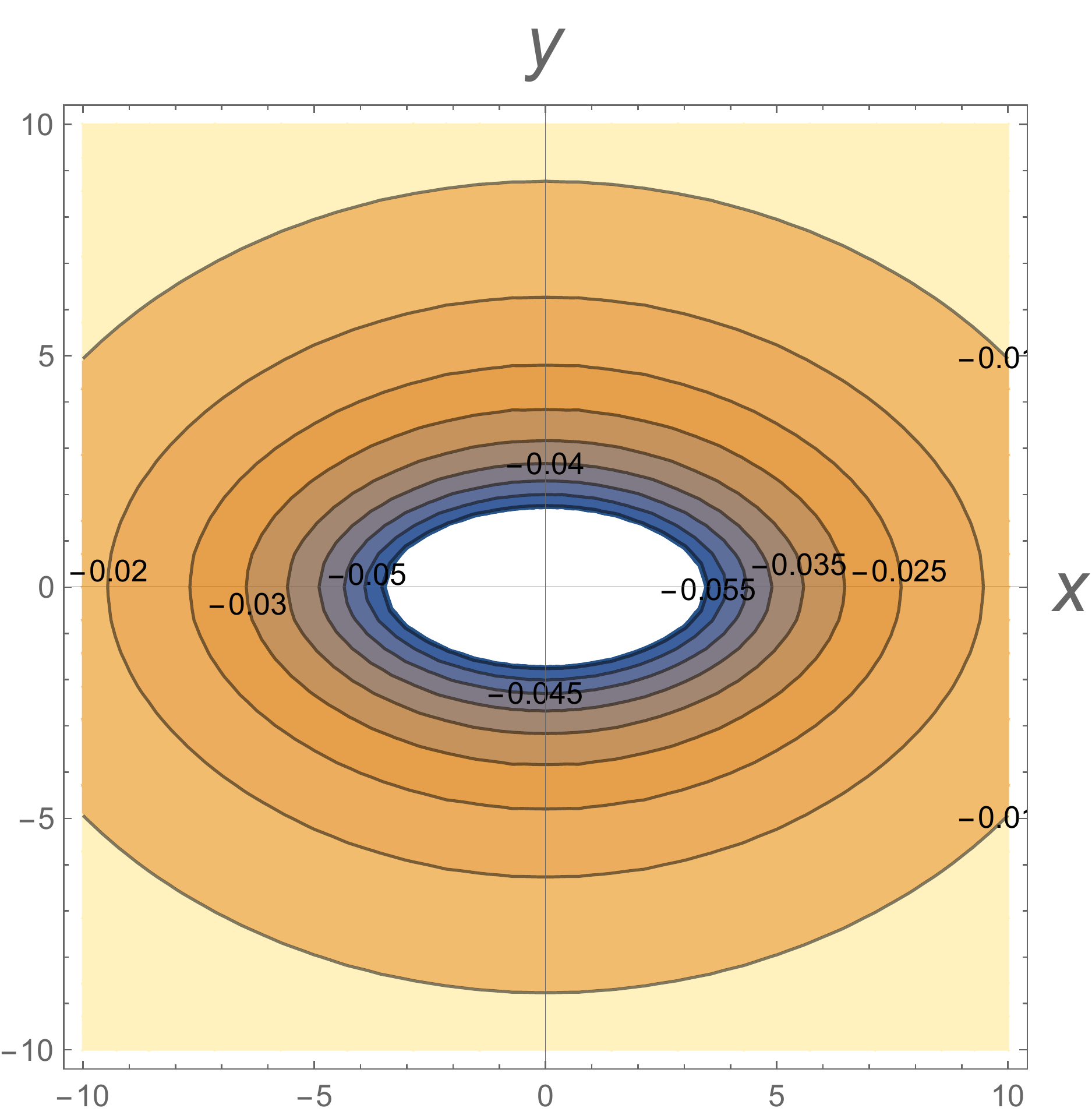}
\caption{ Contours of  $\varphi(X,Y)=\,\,$const.   $X,Y$ are in units of $\Lambda$  for $\gamma=3$.}
\label{f10}
\end{figure}
    \begin{figure}[htb]
\includegraphics[width=7.5cm] {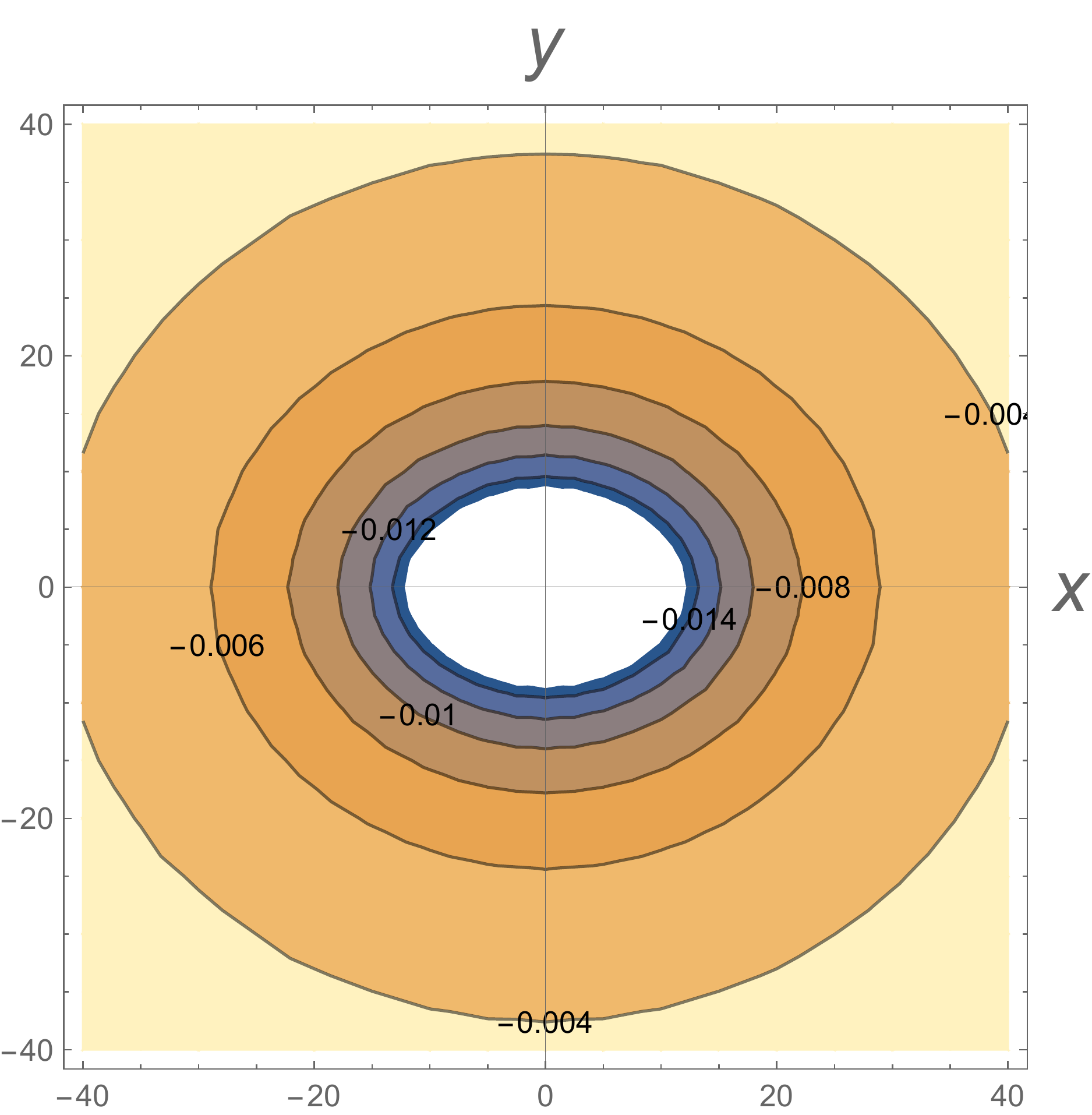}
\caption{ Contours of  $\varphi(X,Y)=\,\,$const at larger distances.   $X,Y$ are in units of $\Lambda$  for $\gamma=3$.}
\label{f11}
\end{figure}

The stream lines of current are given by $(\bm g\times d{\bm L})_z =g_xdy-g_ydx=0$, $d\bm L=(dx,dy)$ is the line element. This  translates to $\varphi=\,\,$const, contours of which are shown in Figs.\,\ref{f10} and \ref{f11}, these contours   are in fact the current lines. One sees that the currents anisotropy decreases with increasing distance, but--at first sight--noting  in current distributions that might suggest an unusual behavior of $h_z(X,Y)$.

As shown in \cite{BKT}, the interaction energy of a vortex at the origin with another one at $(x,y)$ is $\phi_0\varphi(\bm r) /4\pi$. In fields $H$ applied perpendicular to the film, the distance between vortices in the flux-line lattice is $\propto  \sqrt{\phi_0/H}$. Therefore,  
with increasing applied field $H$,   the  anisotropy of contours $\varphi=\,\,$const increases along with   anisotropy parameter of the vortex lattice. The commonly used assumption that the lattice anisotropy is field independent and fixed by the anisotropy of the penetration depth should be used with care. The same can be said about extracting the anisotropy parameter of the film material from geometry of the flux-line lattice.
  
\section{Discussion}
 
In thin films there is no usual differential relation between the in-plane current and the field $h_z$, the Maxwell equation curl$\bm h = 4\pi \bm j/c$ is replaced by boundary conditions at the film which relate the sheet current $\bm g$ to   tangential fields. The common way to evaluate the $h_z$ is to use the Biot-Savart integral relation between $h_z$ and $\bm g$. In our approach, both fields and currents are expressed in terms of the potential $\varphi$.

  London equations {\it per se} are conditions of minimum of the London energy (magnetic + kinetic) \cite{deGennes}. The $h_z(x,y)$ solutions of these equations, however strange they may look,  correspond to the minimum energy. Therefore,  the fact that the vortex magnetic field lines in anisotropic films may prefer to cross the film from the upper half-space to the lower one in some parts of the $xy$ plane, unlike the case of isotropic films where these lines go to infinity without crossing the film again, should be  seen as a way the vortex system minimizes its energy. 
It is worth noting that this situation emerges when the current anisotropy decreases with the distance from the vortex core as shown in Figs.\,\ref{f10} and \ref{f11}. 

It is instructive to consider  an  example of the field $h_z$ at the film created by two concentric current loops in $xy$ plane, the small one   strongly elongated in the $x$ direction  with a large current and the big circular loop with a small current. Both currents are in the same, say, counterclockwise direction so that their contributions to $h_z$ inside a small loop are both positive. However, outside the smaller loop  in its vicinity the contribution of the large current of this loop is negative and cannot be cancelled by positive contribution of the distant loop. This cancellation cannot happen also because the large loop is a circle whereas the small one is elliptic. One can also give a qualitative argument why the domain of negative field is situated near the $y$ axis.  

Hence, although the ``strange" field distribution of Figs.\,\ref{f2}-\ref{f5}  is the consequence of Maxwell and London equations and as such do not bring in any new physics, our result is relevant for  interpretation of data when the normal field component above the film surface is measured such as the Scanning SQUID or Magnetic Force Microscopies. These techniques are currently fast improving, see, e.g. \cite{Zeldov}, and confirming the sign change of the normal field component may become feasible.

Most of our calculation were done for orthorhombic materials with the in-plane anisotropy parameter $\gamma = 3$ and the vortex along $c$. Such materials in fact exist,   examples are NiBi films \cite{NiBi}, or Ta$_4$Pd$_3$Te$_{16}$ \cite{17}.

 \section{Acknowledgements}
The work of V.K. was supported by the U.S. Department of Energy (DOE), Office of Science, Basic Energy Sciences, Materials Science and Engineering Division.  Ames Laboratory  is operated for the U.S. DOE by Iowa State University under contract \# DE-AC02-07CH11358. \\

\appendix
\section{  $\bm{H_{z}(R,Z)}$ in isotropic films}


Rewrite Eq.\,(\ref{h(r,z)}) using $\Lambda$ as a unit length and
$\phi_{0}/2\pi\Lambda^{2}$ as a unit of field: 
\begin{eqnarray}
H_{z}(\bm{R},Z)=h_{z}(\bm{r},z)\frac{2\pi\Lambda^{2}}{\phi_{0}}=\frac{1}{2\pi}\int\frac{d^{2}\bm{q}\,e^{i\bm{q}\bm{R}-qZ}}{1+q},\label{h(R,Z)a}
\end{eqnarray}
where the dimensionless $\bm{q}=\bm{k}\,\Lambda$, $\bm{R}=\bm{r}\,\Lambda$,
and $Z=z/\Lambda$. With the help of identity 
\begin{eqnarray}
\frac{1}{1+q}=\int_{0}^{\infty}e^{-u(1+q)}du\,,\label{ident}
\end{eqnarray}
one rewrites the field as 
\begin{eqnarray}
H_{z}(\bm{R},Z)=\frac{1}{2\pi}\int_{0}^{\infty}du\,e^{-u}\int d^{2}\bm{q}\,e^{i\bm{q}\cdot\bm{R}-q(u+Z)}.\qquad\label{Hz}
\end{eqnarray}
To evaluate the last integral over $\bm{q}$, we note that the three-dimensional
(3D) Coulomb potential is 
\begin{eqnarray}
\frac{1}{4\pi{\cal R}}=\frac{1}{(2\pi)^{3}}\int\frac{d^{3}\bm{Q}}{Q^{2}}\,e^{i\bm{Q}\cdot{\cal R}},\qquad\label{G(R)}
\end{eqnarray}
where $\bm{Q}=(\bm{q},Q_{z})$, ${\cal R}=(\bm{R},Z)$, and $d^{3}\bm{Q}=d^{2}\bm{q}\,dq_{z}$.
Integration over $q_{z}$ is doable: 
\begin{eqnarray}
\int_{-\infty}^{\infty}\frac{dq_{z}}{2\pi}\frac{e^{iq_{z}Z}}{q^{2}+q_{z}^{2}}=\frac{e^{-q|Z|}}{2\,q}\,,\label{Gc2D}
\end{eqnarray}
and we obtain instead of Eq.\,(\ref{G(R)}): 
\begin{eqnarray}
\frac{1}{\sqrt{R^{2}+Z^{2}}}=\frac{1}{2\pi}\int\frac{d^{2}\bm{q}}{q}\,e^{i\bm{q}\cdot R-qZ}.\qquad\label{G1}
\end{eqnarray}
Applying $\partial_{Z}$ to both sides, we obtain: 
\begin{eqnarray}
\int d^{2}\bm{q}\,e^{i\bm{q}\cdot\bm{R}-qZ}=\frac{2\pi Z}{(R^{2}+Z^{2})^{3/2}}.\qquad
\end{eqnarray}
To get the integral in Eq.\,(\ref{Hz}), replace $Z\to Z+u$: 
\begin{eqnarray}
H_{z}(\bm{R},Z)=\int_{0}^{\infty}du\frac{(Z+u)\,e^{-u}}{[R^{2}+(Z+u)^{2}]^{3/2}}\,.\label{hz(r)d}
\end{eqnarray}



\section{  $\bm{H_{z}(\bm{R},Z)}$ in anisotropic films}


Rewrite Eq.\,(\ref{h(R,Z)}), using $\Lambda$ as a unit
length and $\phi_{0}/2\pi\Lambda^{2}$ as a unit of field. Then, transform 
the denominator employing Eq.\,(\ref{ident}): 
\begin{eqnarray}
&&H_{z}(\bm{R},Z) = h_{z}(\bm{r},z)\frac{2\pi\Lambda^{2}}{\phi_{0}}=\frac{1}{2\pi}\int\frac{d^{2}\bm{q}\,q\,e^{i\bm{q}\bm{R}-qZ}}{q+\gamma q_{x}^{2}+q_{y}^{2}/\gamma}\nonumber \\
 &&= 2\pi\int_{0}^{\infty}du\int\frac{d^{2}\bm{q}\,q}{(2\pi)^{2}}e^{i\bm{q}\bm{R}-q\left(u+Z\right)-u(\gamma q_{x}^{2}+q_{y}^{2}/\gamma)}.\qquad\label{h(R,Z)-2}
\end{eqnarray}
Now we use the identity Eq.\,(\ref{Gc2D}) in the form
\begin{eqnarray}
\int_{0}^{\infty}d\xi\int_{-\infty}^{\infty}\frac{dq_{z}}{2\pi}e^{iq_{z}Z-\xi\left(q^{2}+q_{z}^{2}\right)}=\frac{e^{-q|Z|}}{2\,q}.\qquad\label{Gc2D-1}
\end{eqnarray}
\\
Applying $\partial_{Z}$ twice and replacing $|Z|\rightarrow u+Z$,
we obtain
\begin{eqnarray}
q\,e^{-q\left(u+Z\right)}=-2\int_{0}^{\infty}d\xi\int_{-\infty}^{\infty}\frac{dq_{z}}{2\pi}\,q_{z}^{2}\,e^{iq_{z}\left(u+Z\right)-\xi(q^{2}+q_{z}^{2})}.\qquad\label{ident-1}
\end{eqnarray}
Hence, we have 
\begin{widetext}
\begin{eqnarray}
H_{z}(\bm{R},Z) & = & -4\pi\int_{0}^{\infty}d\xi\int_{0}^{\infty}du\int\frac{d^{3}\bm{Q}}{(2\pi)^{3}}q_{z}^{2}\exp\left[i\bm{q}\bm{R}+iq_{z}\left(u+Z\right)-u(\gamma q_{x}^{2}+q_{y}^{2}/\gamma)-\xi(q^{2}+q_{z}^{2})\right],\quad\bm{Q}=(\bm{q},q_{z}).\qquad\label{h(R,0)}
\end{eqnarray}
\end{widetext}

The required expressions, Eq.\,(\ref{Hz1short}) and Eq.\,(\ref{Hz1lift}),
can be obtained in the following tedius but straightforward procedure:
(1) Change the variables as $\xi\rightarrow\eta$ according to $\xi=u/\eta$,
(2) perform the Gaussian integrations over $Q$, and then (3) integrate
over $u$, leaving the remaining $\eta$ integration to be performed
numerically. In particular, the $u$ integration can be performed
analytically according to the formula 
\begin{equation}
\int_{0}^{\infty}duu^{\nu-1}\exp\left(-\alpha^{2}u-\beta^{2}/u\right)=2\left(\beta/\alpha\right)^{\nu}K_{\nu}\left(2\alpha\beta\right),
\end{equation}
where $K_{\nu}\left(z\right)$ is the modified Bessel function of
the second kind \cite{Abramowitz}, e.g. 
\begin{equation}
K_{1/2}\left(z\right)=\sqrt{\pi/2z}e^{-z},\:K_{3/2}\left(z\right)=\sqrt{\pi/2z}\left(1+z^{-1}\right)e^{-z}.
\end{equation}


\end{document}